\documentclass[letter]{aa} 

\usepackage{graphicx}
\usepackage{txfonts}
\usepackage{hyperref}
\hypersetup{
    colorlinks=true,
    linkcolor=blue,
    urlcolor=magenta,
    citecolor=blue
    }
\usepackage{array}
\usepackage{multirow} 
\usepackage{float} 
\usepackage{multicol}

\begin{document}

   \title{Detecting H$_2$O with CRIRES+: the case of WASP-20b\thanks{\scriptsize{Based on observations collected at the European Southern Observatory under ESO programme 107.22SX.001.}} }


   \author{M.~C. Maimone
          \inst{1,2}
          \and
          M. Brogi 
          \inst{3,4,5}
          \and
          A. Chiavassa
          \inst{1,6}
          \and
          M.~E. van den Ancker 
          \inst{2}
          \and
          C.~F. Manara
          \inst{2}
          \and
          J. Leconte
          \inst{7}
          \and
          S. Gandhi 
          \inst{8,3,5}
          \and
          W. Pluriel 
          \inst{9}
          }

  \institute{Universit\'e C\^ote d'Azur, Observatoire de la C\^ote d'Azur, CNRS, Lagrange, CS 34229, Nice,  France \\
                \email{mmaimone@oca.eu}
                \and
                 European Southern Observatory, Karl-Schwarzschild-Str. 2, 85748 Garching, Germany
                 \and
                 Department of Physics, University of Warwick, Coventry CV4 7AL, UK 
                 \and
                 INAF–Osservatorio Astrofisico di Torino, Via Osservatorio 20, I-10025, Pino Torinese, Italy 
                 \and
                 Centre for Exoplanets and Habitability, University of Warwick, Gibbet Hill Road, Coventry CV4 7AL, UK
                 \and
                 Max-Planck-Institut f\"{u}r Astrophysik, Karl-Schwarzschild-Stra\ss{}e 1, 85741 Garching, Germany
                \and
                Laboratoire d'astrophysique de Bordeaux, Univ. Bordeaux, CNRS, B18N, all\'ee Geoffroy Saint-Hilaire, 33615 Pessac, France
                \and
                Leiden Observatory, Leiden University, Postbus 9513, 2300 RA Leiden, The Netherlands
                \and
                Département d’astronomie de l’Université de Genève, Chemin Pegasi 51, 1290 Versoix, Switzerland
             }

   \date{Received; accepted}

 
  \abstract
   {Infrared spectroscopy over a wide spectral range and at the highest resolving powers (R$>$70\, 000) has proved to be one of the leading technique to unveil the atmospheric composition of dozens of exoplanets. The recently upgraded spectrograph CRIRES instrument at the VLT (CRIRES+) was operative for a first Science Verification in September 2021 and its new capabilities in atmospheric characterisation were ready to be tested.
   }
   {We analysed transmission spectra of the Hot Saturn WASP-20b in the K-band (1981-2394 nm) acquired with CRIRES+, aiming to detect the signature of H$_2$O and CO.}
   {We used Principal Component Analysis to remove the dominant time-dependent contaminating sources such as telluric bands and the stellar spectrum and we extracted the planet spectrum by cross-correlating observations with 1D and 3D synthetic spectra, with no circulation included.   
   }
   {We present the tentative detection of molecular absorption from water-vapour at S/N equal to 4.2 and 4.7 by using only-H$_2$O 1D and 3D models, respectively.  The peak of the CCF occurred at the same rest-frame velocity for both model types (V$_{\rm{rest}}$=$-1\pm{1}$ km s$^{-1}$), and at the same projected planet orbital velocity but with different error bands (1D model: K$_{\rm{P}}$=131$^{+18}_{-29}$  km s$^{-1}$; 3D: K$_{\rm{P}}$=131$^{+23}_{-39}$ km s$^{-1}$). Our results are in agreement with the one expected in literature (132.9 $\pm$ 2.7 km s$^{-1}$).
    
   }
   {Although sub-optimal observational conditions and issues with pipeline in calibrating and reducing our raw data set, we obtained the first tentative detection of water in the atmosphere of WASP-20b.  We suggest a deeper analysis and additional observations to confirm our results and unveil the presence of CO. } 

   \keywords{Planets and satellites: atmospheres -- Planets and satellites: individual (WASP20b) -- Techniques: spectroscopic -- methods: data analysis }

   \maketitle
%

\section{Introduction}

The remote atmospheric characterisation of exoplanets is a key milestone to unveil their physical and chemical processes (\citealt{2009IAUS..253..263M}), their formation history (\citealt{2014AAS...22320702M}, \citealt{2018IAUS..332...69E}), and ultimately, the presence of conditions suitable for life  (\citealt{2016PhDT.......143S}).
In recent years, High-Resolution Spectroscopy (R > 25\,000) has become a tool at the forefront for acquiring exoplanets' spectra. At high resolution,  molecular bands are resolved into a forest of individual lines, which enables a line-by-line comparison with synthetic spectra through cross-correlation and allows to disentangle the planetary signal -- Doppler-shifted during the transit -- from the stationary or quasi-stationary signals, like telluric bands and stellar spectral lines (\citealt{2010Natur.465.1049S}). 

The high demand in terms of signal-to-noise for the characterisation of exoplanets has always been the major limit of this technique, but in the last decades a new generation of ground-based instruments have been built to fulfil the minimum  requirements in terms of resolution, stability, and adequate-collective area to secure solid detections in the near-infrared.
At infrared wavelengths, VLT/CRIRES has been the first successful instrument, obtaining the first detection of CO in transmission (\citealt{2010Natur.465.1049S}) and the first detection of H$_2$O in emission (\citealt{2013MNRAS.436L..35B}). Water was then confirmed with Keck/NIRSPEC by \cite{2014ApJ...783L..29L}, while \cite{2014A&A...565A.124B} detected CO and H$_2$O simultaneously, and \cite{2016ApJ...817..106B}  provided a first measurement of winds and rotation, both results obtained with CRIRES. After CRIRES was decommissioned, other instruments were successful, starting with the detection of TiO via Subaru/HDS (\citealt{2017AJ....154..221N}). \cite{2018A&A...615A..16B} presented the first detection of CO+H$_2$O with TNG/GIANO, while \cite{2019A&A...621A..74A} found water in the J and Y bands of CAHA/CARMENES. In 2021, CHFT/SPIRou also presented detections of H$_2$O (\citealt{2021AJ....162..233B}) and CO (\citealt{2021AJ....162...73P}). In the meantime, \cite{2021Natur.592..205G}  presented the simultaneous detection of 6 molecular species with GIANO and \cite{2021Natur.598..580L} achieved the most precise abundance measurement to date with Gemini-S/IGRINS.
Recently, the CRIRES instrument has been upgraded into a cross-dispersed spectrograph (CRIRES+, \citealt{2016SPIE.9908E..0ID}) and started operating in September 2021. CRIRES+ allows high sensitivity in the infrared range (0.95-5.3 $\mu$m), where two main carriers of carbon (CO) and oxygen (H$_2$O) simultaneously imprint the spectrum (\citealt{2012EGUGA..1413720M}). 
Using CRIRES+, we observed WASP-20b during the Science Verification (SV) time of the instrument. Our goal was a demonstration of the basic capabilities of the new instrument and we chose to observe WASP-20 because it was the only target with a visible transit during the SV observing window. \\ WASP-20b is an  in a 4.9-day, near-aligned orbit around a F9-type star and with an equilibrium temperature of 1379K. It was observed for the first time by \cite{2015A&A...575A..61A}, discovered as a binary system separated by only 0.26" by \cite{2016ApJ...833L..19E} and confirmed by \cite{2020A&A...635A..74S}.  \\
In the following, we present the analysis of transmission spectra acquired with CRIRES+ which led us to the first, tentative detection of water-vapour in the atmosphere of WASP-20b.



\section{Observations and Data Reduction}\label{Sec:Obs}

\begin{table}
\footnotesize
\begin{center}
\caption{Overview of WASP-20b observations during the first night of the Science Verification run of CRIRES+, as reported by the Paranal Differential Image Motion Monitor (DIMM).
}\label{tableobs}
 \begin{tabular} {|l|l|}
 \hline
Programme ID & 107.22SX.001 \\
Night & 2021-09-16;  1:40UT - 6:30UT \\
Phase & 0.98 - 0.02 \\
N$_{obs}$ & 75 (63 + 12)\\
Exp. Time & 1 x 180s\\
Obs. Mode & Nodding \scriptsize{A-A-B-B-B-B-A-A}\\
Slit & 0.2" \\
AO loop & Closed \\
Max Resolution  & 92\,000 \\
Wavelength Setting & K2217  (1981-2394 nm) \\
Airmass & 1.006-1.455 \\
S/N & 19 - 38 (AVG=28)\\
Seeing (towards target) & 0.82"-1.27"  \\
\hline
\end{tabular}
\end{center}
 \tablefoot{N$_{obs}$ is the total number of observed spectra, acquired before (63) and after (12) crossing the Zenith avoidance area. The exposure time is expressed as NDIT × DIT, where DIT is the detector integration time and NDIT is the number of detector integrations.}
\end{table}

We observed WASP-20b for 5 hours before, during, and after the 3.4-hour transit occurred on Sept, 16 2021. During the night, the target crossed the Zenith avoidance area of the telescope, causing a gap of 1 hour in the acquisition. An overview of observations is shown in Table \ref{tableobs}. We set the spectrograph to cover the wavelength range 1981-2394 nm (K-2217 band), roughly centered on the branch of 2-0 Ro-Vibrational transitions of carbon monoxide and where additional molecular absorption from water-vapour, carbon monoxide and methane is possible (\citealt{2020MNRAS.495..224G}). We employed CRIRES+ at the maximum resolution (R = 92, 000\footnote{\scriptsize{Documented in the CRIRES user manual available at \url{https://www.eso.org/sci/facilities/paranal/instruments/crires/doc.html}}}) by using the 0.2" slit. Target and sky spectra were taken with the nodding acquisition mode AABBBBAA for background subtraction. 
Although the airmass remained acceptable during all the night (1.0 - 1.45) and the time-averaged seeing in the direction of the target did not overcome 1.27", we obtained an averaged signal-to-noise (S/N) of only 28. 
This value agrees with the one found by \cite{2022arXiv220610621H}, but it turned out to be low if compared to other high-resolution observations (see Sec. \ref{Sec:DiscCon} for details). We expected this to impact negatively on our final result.  

According to results from \cite{2016ApJ...833L..19E} and \cite{2020A&A...635A..74S}, we also checked for traces of binarity by looking at vertical slices of our raw observations. We found no significantly resolved double peaks that could explain the presence of a second star. We therefore treated WASP-20 as a single star.

We performed the calibration and extraction of the spectra with the CRIRES+ pipeline (version 1.0.4, \citealt{valenti_rodler_brulacassi}), run through the esoreflex workflow (version 2.11.3, \citealt{2013A&A...559A..96F}) and the command-line interface esorex (version 3.13.5)\footnote{\scriptsize{Documentation available at the ESO website \url{http://www.eso.org/sci/software/pipelines/}}}. We used calibration frames that were acquired at the beginning of the night and just after the interruption for flat-fielding, dark-correction, wavelength calibration and to take into account the non-linearity between pixels and wavelengths. CRIRES+ has been equipped with 3 detectors (henceforth CHIP1, CHIP2, CHIP3), each of them divided into eight orders. We successfully reduced and extracted raw spectra from orders 3,4,5,6,7 of CHIP1 and orders 3,4,5,6,7,8 of CHIP2. The data reduction of all orders of CHIP3 and the 8th order of CHIP1 failed because the pipeline could not find the pixel-wavelength conversion tables (TraceWave Tables) associated to those orders. Each full AABBBBAA nodding sequence (8 exposures) was combined at the level of the pipeline into a single reduced spectrum. At the end of the data reduction, we obtained 18 reduced spectra, 6 of them out-of-transit and 12 in-transit. For our analysis, we used only in-transit spectra. 

After the data reduction, we noticed that the wavelength calibration performed with the pipeline was not accurate. In particular, the location of spectral lines at the edges of each order were clearly shifted further than spectral lines near the center of each order. We attempted to use our custom wavelength re-calibration pipeline designed for the old CRIRES (\citealt{2019A&A...631A.100C}) to re-aligned spectra. The code compares a telluric model with the data, and searches for a wavelength solution through  Monte Carlo Markov Chains. The higher the number of telluric absorption lines in a spectral range, the easier the comparison between the model and the data. Unfortunately, the calibration failed because of the lack of absorption lines within 2114-2140 nm and 2210-2222 nm.
We had two options at this point: (1) using Principal Component Analysis (PCA) with not-aligned spectra and preserve the original S/N, dealing with a possible stellar contamination; (2) removing those problematic orders (a total of 24 spectra), apply the technique developed by \cite{2019A&A...631A.100C} to subtract stellar signal through 3D models and reduce stellar contamination before PCA, but risking to have insufficient planetary signal for a detection. We decided to proceed with the first option.

\begin{table}
\footnotesize
\begin{center}
\caption{System parameters from \cite{2015A&A...575A..61A} compared to the ones used in this work as inputs for 1D and 3D models.}\label{tablemodels} 
 \begin{tabular} {m{1.3cm} m{1.cm} m{1.cm}}
\hline
\multicolumn{1}{l}{\multirow{2}{*}{\textbf{Parameters}}} 
     & \multicolumn{1}{|c}{\multirow{2}{*}{\textbf{Anderson et al. 2015}}}  & \multicolumn{1}{|c}{\textbf{Inputs Models}}  \\
  &\multicolumn{1}{|c}{}  & \multicolumn{1}{|c}{\textbf{in This Work}} \\
 \hline
 M$_P$ (M$_J$) & \multicolumn{1}{c}{0.311 $\pm$ 0.017} &   \multicolumn{1}{c}{0.311}  \\ [.5ex]
 R$_P$ (R$_J$) & \multicolumn{1}{c}{1.462 $\pm$ 0.059}   &  \multicolumn{1}{c}{1.462}    \\ [.5ex]
 T$_{eq}$ (K)  & \multicolumn{1}{c}{1379 $\pm$ 31} &  \multicolumn{1}{c}{1400}   \\ [.5ex]
 g (m/s$^2$)  & \multicolumn{1}{c}{3.36 $\pm$ 0.28} &  \multicolumn{1}{c}{3.77}   \\ [.5ex]
 \multirow{2}{*}{p (bar)} & \multicolumn{1}{c}{\multirow{2}{*}{--}} & \multicolumn{1}{c}{$10^2 - 10^{-8}$ (1D) }    \\ 
 & & \multicolumn{1}{c}{2x$10^2 - 10^{-9}$ (3D)}  \\ [.5ex]
  
 \hline
 \end{tabular}
\end{center}
\label{tablemodels}

\end{table}

\section{Methodology: PCA and Cross-Correlation}\label{Sec:Methodology}

At this initial stage of the analysis, ground based high-resolution observations are dominated by telluric bands and the stellar spectrum, which are orders of magnitude stronger than the exoplanet signals of interest and therefore can be considered as contaminant signals in the study of exoplanetary atmospheres. To extract the planet’s spectrum, we executed  the two following steps as done in \cite{2021Natur.592..205G}.

\begin{enumerate}
    \item \textbf{Principal component analysis} (\textbf{PCA}). The PCA can identify and remove the dominant time dependent contaminating sources that are quasi-stationary in wavelength, such as telluric bands (black vertical lines in Fig.~\ref{PCAremoval}a), stellar lines, and  systematic trends caused by instruments, and leave the planet signal and the uncorrelated noise as residuals (Fig.~\ref{PCAremoval}b). Typically,  the number of removed components varies between 2 and 8, depending on the quality of the data (\citealt{2021Natur.592..205G}). Before applying the PCA, we corrected  the raw spectra for detector cosmetics and cosmic rays by substituting bad-pixels with NaN strings. This made easier to mask bad-pixels afterwards in the analysis. \\ The spectra were normalised by their median value to correct for changes in the overall amount of flux that reaches detectors due to variable transparency, imperfect telescope pointing, or instability of the stellar point spread function. Subsequently, each spectral channel (each column) had its mean subtracted and each spectrum (each row) was divided by its standard deviation and given as input to the Singular Value Decomposition (SVD) Pyhton function numpy.linalg.svd\footnote{\scriptsize{Available at \url{https://numpy.org/doc/stable/reference/ generated/numpy.linalg.svd.html}}}. The output was a matrix of eigenvalues, extracted for a given number of components. We performed a multi-linear regression between the eigenvalues found via SVD and the matrix of fluxes after median normalisation, and we divided the latter by the resulting fit. Lastly, a high-pass gaussian filter with a FWHM of 80 pixels was applied to the data and residual outliers were masked. \\
   
    \begin{figure}
   \centering
    \begin{tabular}{c}
     \includegraphics[width=0.99\hsize]{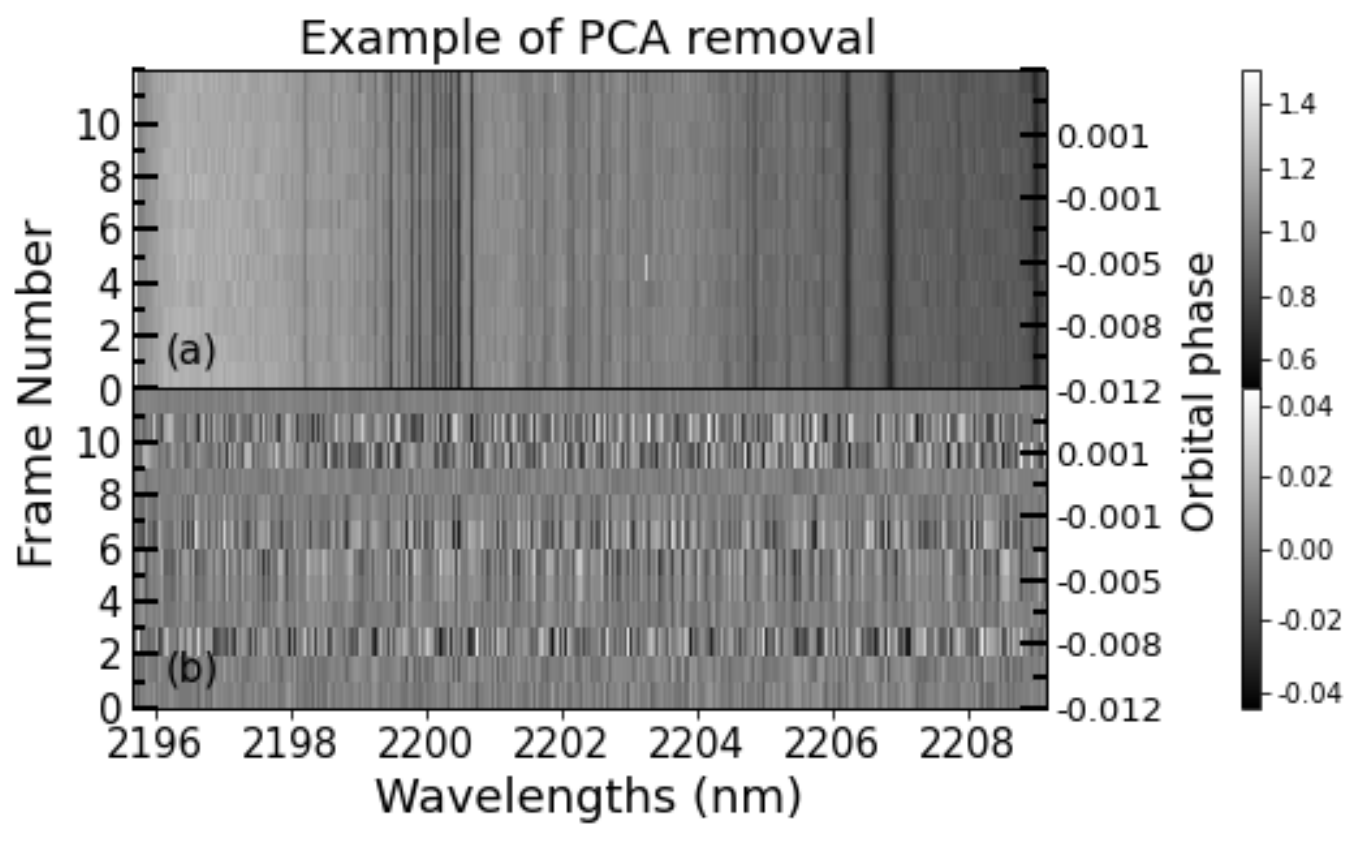}
     
      \end{tabular}
      \caption{Example of PCA removal on WASP-20 data, observed during the first night of the Science Verification time with CRIRES+. The sequence of 12 normalised spectra is shown in the wavelength range 2195-2209 nm (5th order of CHIP1) before (panel a) and after (panel b) the removal of the first 7 principal components.}
        \label{PCAremoval}
\end{figure}
    
    \item \textbf{Cross-Correlation} of WASP-20b data with transmission synthetic spectra to extract the planet signal. Indeed, even if a single absorption line has a S/N $\ll$ 1, there are hundreds of strong molecular lines in the CRIRES+ K-band. By co-adding them into a single cross-correlation function (CCF),  the planet’s faint signal is enhanced by a factor of approximately $\sqrt{N_{lines}}$ (\citealt{2014A&A...565A.124B}), allowing us to attempt a detection of the planet signature. \\

    
    Because with CRIRES+ we did not resolve the binary nature of WASP-20 found by \cite{2016ApJ...833L..19E} and \cite{2020A&A...635A..74S}, we treated the system as a single star (as noted in Sec. \ref{Sec:Obs}) and took as main reference \cite{2015A&A...575A..61A}. The final result is strictly related to this choice: different planetary mass and radius reflect on differences in surface gravity and atmospheric pressure, which have direct consequences on the atmospheric molecular absorption and on the shape of the planetary spectrum that we cross-correlated with data. 
    
    In this work, we used synthetic spectra computed from 1D models using GENESIS (\citealt{2017MNRAS.472.2334G}), and from SPARC/MIT Global Circulation Models (GCM, \citealt{2013ApJ...762...24S}, \citealt{2021MNRAS.501...78P}, \citealt{2022A&A...658A..42P}) using Pytmosph3R (\citealt{2019A&A...623A.161C}, \citealt{2022A&A...658A..41F}).
    We selected the closest GCMs in terms of mass, radius and temperature to WASP-20b (see Table~\ref{tablemodels}) and we used the same values to calculate 1D models. Only the pressure ranges were slightly different. We chose 3D GCMs that did not include dynamics (i.e., planet rotation, circulation and winds) to make the comparison  1D/3D as plain as possible. In this way, the changes in line shape and intensity are ascribable only to the inherent three-dimensional structure of the planet T-p profile. 
    All models were computed with an isothermal profile of 1400K, with no thermal inversion nor clouds.  We firstly assumed a solar metallicity and we computed abundances at chemical equilibrium for all the species, then we included only CO and H$_2$O in the final models used in our analysis.
    As shown by \cite{2020MNRAS.495..224G}, at 1400K and at 2.3 $\mu$m,  CO and H$_2$O are the dominant sources of opacities, hence we computed three synthetic spectra per each model type, containing: (i) only CO molecules, (ii) only H$_2$O, and (iii) CO+H$_2$O (Fig.\ref{Modelfullrange}). CO opacities of GCMs and H$_2$O opacities of both model types were taken from the ExoMol database (\citealt{2018MNRAS.480.2597P}), while CO opacities of 1D models where taken from the HITEMP database (\citealt{2015ApJS..216...15L}). We assumed the same volume mixing ratio (VMR) for water-vapour in  both model type (log[H$_2$O] = -3.3 ), whereas for CO the VMR was slightly different (log[CO] = -3.4 and -3.35 in 1D and 3D models, respectively). A summary is given in Table \ref{tablemodels} and in Fig.~\ref{Modelfullrange}.
\end{enumerate}

    %
    

\begin{figure*}[th!] 
	\centering
	     
	     \includegraphics[width=1.\hsize]{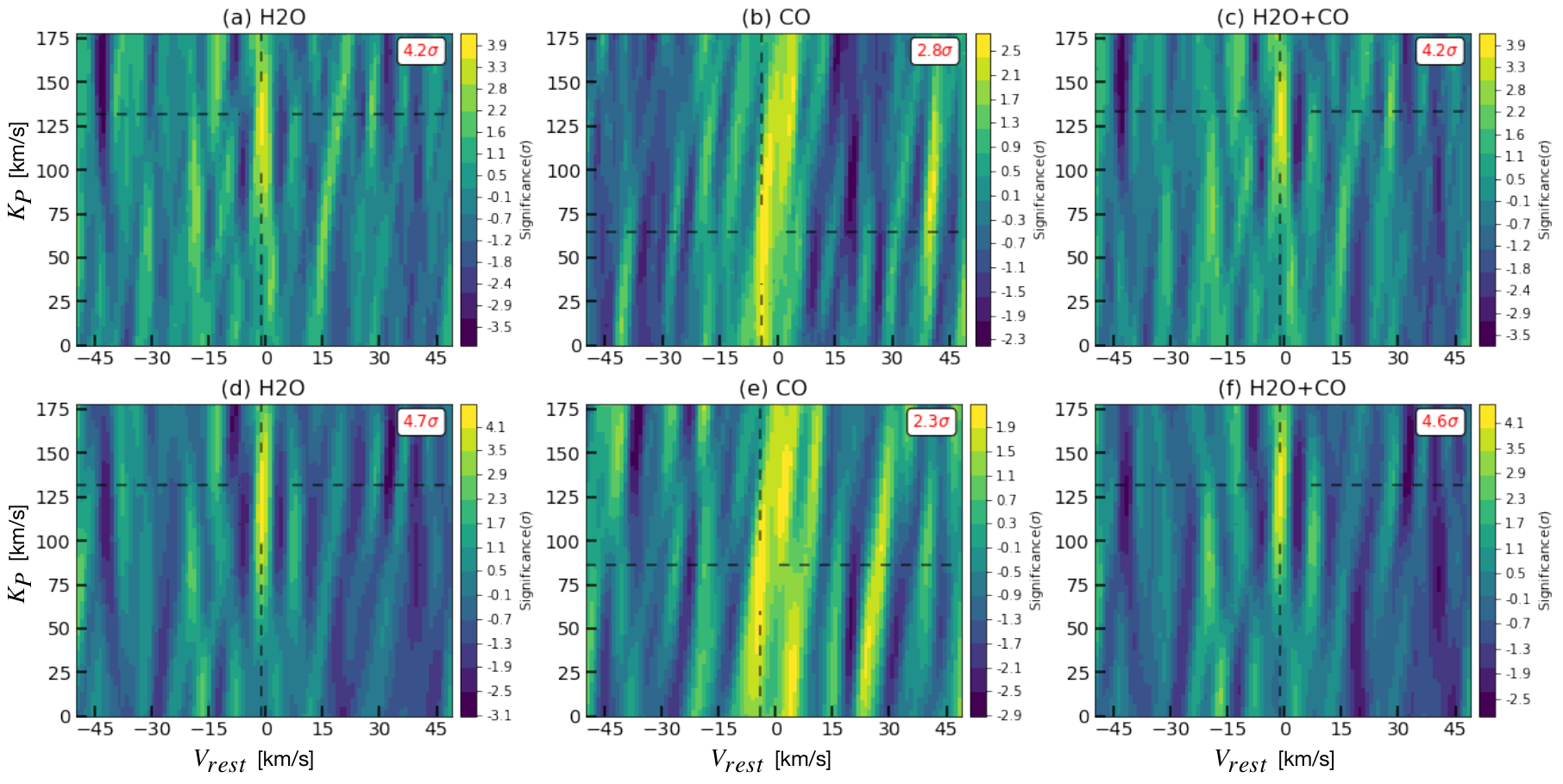} 
		 
	 \caption{
      Total cross correlation signal from only-H$_2$O (panels a and d), only-CO ( panels b and e) and H$_2$O+CO (panels c and f) for the atmosphere of WASP-20b, shown as a function of rest-frame velocity and planet projected orbital velocity. 1D models were used in the top row, whereas 3D GCM models in the bottom one. Both model type are cloud-free, in chemical equilibrium and with no thermal inversion and contains a VMR equals to $-3.3$ for H$_2$O and to $-3.4$ and $-3.5$ for CO in 1D and 3D models, respectively. 
      }\label{fig:Results}

\end{figure*}

	     
		  
		


\section{Results}\label{Sec:Results}
Following step-by-step the methodology explained in Sec.~\ref{Sec:Methodology}, we obtained the results shown in Fig.~\ref{fig:Results}. The six figures represent the total strength of the cross-correlation signal as a function of the planet rest-frame velocity V$_{\rm{rest}}$ and projected orbital velocity K$_{\rm{P}}$ calculated by using only-H$_2$O (a,d), only-CO (b,e) and H$_2$O+CO (c,f) 1D models (top row) and 3D GCM models (bottom row). \\ Water vapour is detected with both 1D and 3D only-H2O models at a S/N of 4.2 and 4.7, respectively, after removing the first 7 principal components with the PCA. The level of detection was estimated by dividing the peak value of the cross-correlation by the standard deviation of the noise. The peak occurred in correspondence of the same velocities for both the model types, but with different error bars for K$_{\rm{P}}$: V$_{\rm{rest}}$=$-1\pm{1}$ km s$^{-1}$, K$_{\rm{P}}$=131$^{+18}_{-29}$  km s$^{-1}$ with 1D model and K$_{\rm{P}}$=131$^{+23}_{-39}$ km s$^{-1}$ with the 3D model. \\
We compared our results with the RV semi-amplitude expected from \cite{2015A&A...575A..61A} results. We calculated the orbital velocity V$_{\rm{orb_A}}$ = $\frac{2\pi a}{P}$ = (133.3 $\pm$ 1.5) km s$^{-1}$ and obtained a K$_{\rm{P_A}}$ = V$_{\rm{orb_A}}$ sin${i_A}$ = (132.9 $\pm$ 2.7) km s$^{-1}$, which is in agreement with either our 1D and 3D K$_{\rm{P}}$  values. \\
No significant cross-correlation signal was obtained for carbon monoxide (S/N$ < 3$) with the chosen models. The peak seems rather split and spread over a wide range of K$_{\rm{P}}$, with the maximum shifted towards lower K$_{\rm{P}}$  (86 and 64 km s$^{-1}$ with 1D model and GCM, respectively)  and negative  V$_{\rm{rest}}$ ($-4$ km s$^{-1}$ in both cases). It is possible that this is caused by a strong contamination from  CO absorption lines of the star, left in the residuals during the PCA.
The cross-correlation with the H$_2$O+CO models reflects the signature of water vapour, but with a lower S/N level (4.2 and 4.6) due to the pollution of the CO contribution. \\


We determined the statistical significance of the H$_2$O signal as in previous work (\citealt{2012Natur.486..502B}, \citealt{2013ApJ...767...27B}). From the matrix containing the cross correlation signal as function of planet radial velocity and time, CCF(V, t), we selected those values not belonging to the planet RV curve (out-of-trail) and those ones belonging to the planetary trace (in-trail). 
In Fig.~\ref{in-out_trail} we show histograms made of out-of-trail values with a yellow solid line and the ones made of in-trail values with a blue solid line, obtained with the 1D model (left panel) and 3D (right panel) models.  It is clear that the distribution of the cross-correlation noise is centered at 0 and that the distribution containing the planet signal is systematically shifted toward higher values. This can be tested statistically by using a Welch t-test. For the signal from WASP-20b, we were able to reject the hypothesis that the out-of-trail and in-trail distributions are drawn from the same parent distribution at 3.1$\sigma$ level of confidence for the 1D model and at 4.1$\sigma$ for the 3D model (see Fig.~\ref{in-out_trail}).

	     
		  
		


\section{Discussion and Conclusions} \label{Sec:DiscCon}
We reported the first tentative detection of water-vapour in the atmosphere of WASP-20b made through the recently-upgraded spectrograph CRIRES+. \\ No significant peak was found for the carbon monoxide, likely due to a considerable contamination from stellar absorption. \\
Our H$_2$O detection significance (3.1$\sigma$ with the 1D model and 4.1$\sigma$ with 3D model) is slightly lower than the ones achieved with other instruments in literature always in transmission: e.g., \cite{2016ApJ...817..106B} detected H$_2$O in the atmosphere of HD 189733b with a statistical significance equal to 4.8$\sigma$ in our same band with VLT/CRIRES; \cite{2019A&A...621A..74A} and \cite{2019A&A...630A..53S} found water vapour respectively at 7.5$\sigma$ in HD 189733b and and 8.1$\sigma$ in HD 209458b, both around 1.15 and 1.4 $\mu$m with CAHA/CARMENES; also \cite{2021Natur.592..205G} detected water vapour in HD 209458b, but at 9.6$\sigma$ in the range 0.95–2.45 $\mu$m with TNG/GIARPS. 
However, a crucial difference between our observations and the other ones is the initial S/N: e.g., \cite{2021Natur.592..205G} measured a mean S/N between $\sim$80 and $\sim$120 per spectrum per pixel averaged across the entire dataset and the entire spectral range; the typical continuum S/N per spectrum reached in \cite{2019A&A...621A..74A}  was $\sim$150;  the mean S/N for \cite{2019A&A...630A..53S} was of $\sim$85 for the bands at  1.15 $\mu$m and of $\sim$65 for the band at 1.4 $\mu$m in the first half of the observations, even though it dropped respectively below 60 and 50 in the second half.  We could not find information about the S/N of observations in \cite{2016ApJ...817..106B}. On the contrary, our observations reached a mean of only 28 (see Sec. \ref{Sec:Obs} and Table \ref{tableobs}).  \\
Perhaps, a thorough and accurate comparison with other observations could have shed light on the causes of the lower S/N (if, e.g., it was due only to the instrument, only to the particular target choice or a combination of both), but that was beyond the purpose of this work. However, we tried to identify potential weaknesses to be avoided or corrected in the future. For example, a wrong NDIT (NDIT=2 instead of 1) was chosen during the preparation of these observations, causing a halving of temporal resolution. Moreover, during the night the target crossed the zenith avoidance area, resulting in an observational gap of one hour and in a loss of 1/3 of the planet’s transit. Future observations should consider a different period to avoid the gap. In addition, the pipeline that we used failed in reducing more than 1/3 of the dataset (all orders of CHIP3 and the 8th order of CHIP1) and was inaccurate in the wavelength solution. Particularly for the CO, this prevented us to correct the dataset for stellar contamination, which, as done in \cite{2019A&A...623A.161C} and \cite{2019AJ....157..209F}, can increase the significance and allow a clearer detection.

Future deeper analysis should firstly aim at refining the wavelength solution by using the updated version of the pipeline and correcting for the spectrograph instability at the sub-pixel level.
Future observations are equally, strongly suggested to improve our preliminary results. The more efficient ABBA nodding pattern should be preferred to gain in temporal resolution. The S/N of CO of WASP-20b could be noticeably increased by observing at longer wavelengths of CRIRES+ close to, e.g., 4.5 $\mu$m (\citealt{2014A&A...561A.150D}).

\begin{acknowledgements}
      This project has been developed during an ESO Studentship [M.C.M.] and has received funding from the European Research Council (ERC) under the European Union’s Horizon 2020 research and innovation programme (grant agreement n◦ 679030/WHIPLASH [M.C.M.]; and project {\sc Four Aces}, grant agreement No 724427 [W.P.]). It has also been carried out in the frame of the National Centre for Competence in Research PlanetS supported by the Swiss National Science Foundation (SNSF) [W.P.]. W.P. acknowledges financial support from the SNSF for project 200021\_200726. M.B. acknowledges support from the UK Science and Technology Facilities Council (STFC) research grant ST/T000406/1. J.L. also acknowledges funding from the french state: CNES, Programme National de Planétologie (PNP), the ANR (ANR-20-CE49-0009: SOUND).\\ For this work, it was granted access to the HPC resources of Observatoire de la C\^ote  d'Azur $--$ M\'esocentre SIGAMM. This research made use of IPython, Numpy, Matplotlib, SciPy, and Astropy\footnote{Available at \url{http://www.astropy.org/}}, a community-developed core Python package for Astronomy \citep{2013A&A...558A..33A}. 
\end{acknowledgements}

\bibliographystyle{aa}
\bibliography{biblio.bib}

\appendix
\section{ }
\begin{figure*}[th]
   \centering
    \begin{tabular}{cc}
     \includegraphics[width=0.68\hsize]{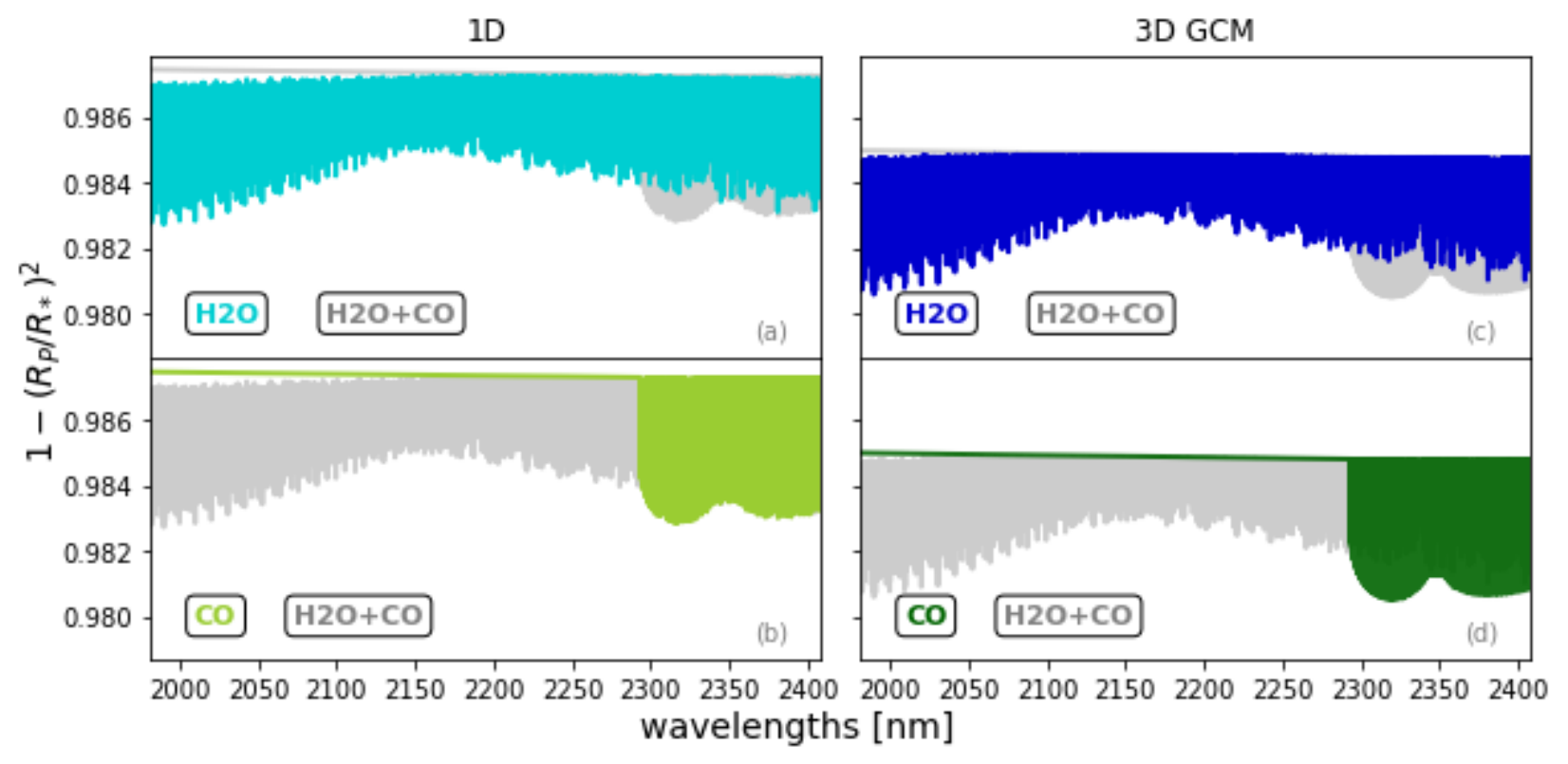}
     \includegraphics[width=0.29\hsize]{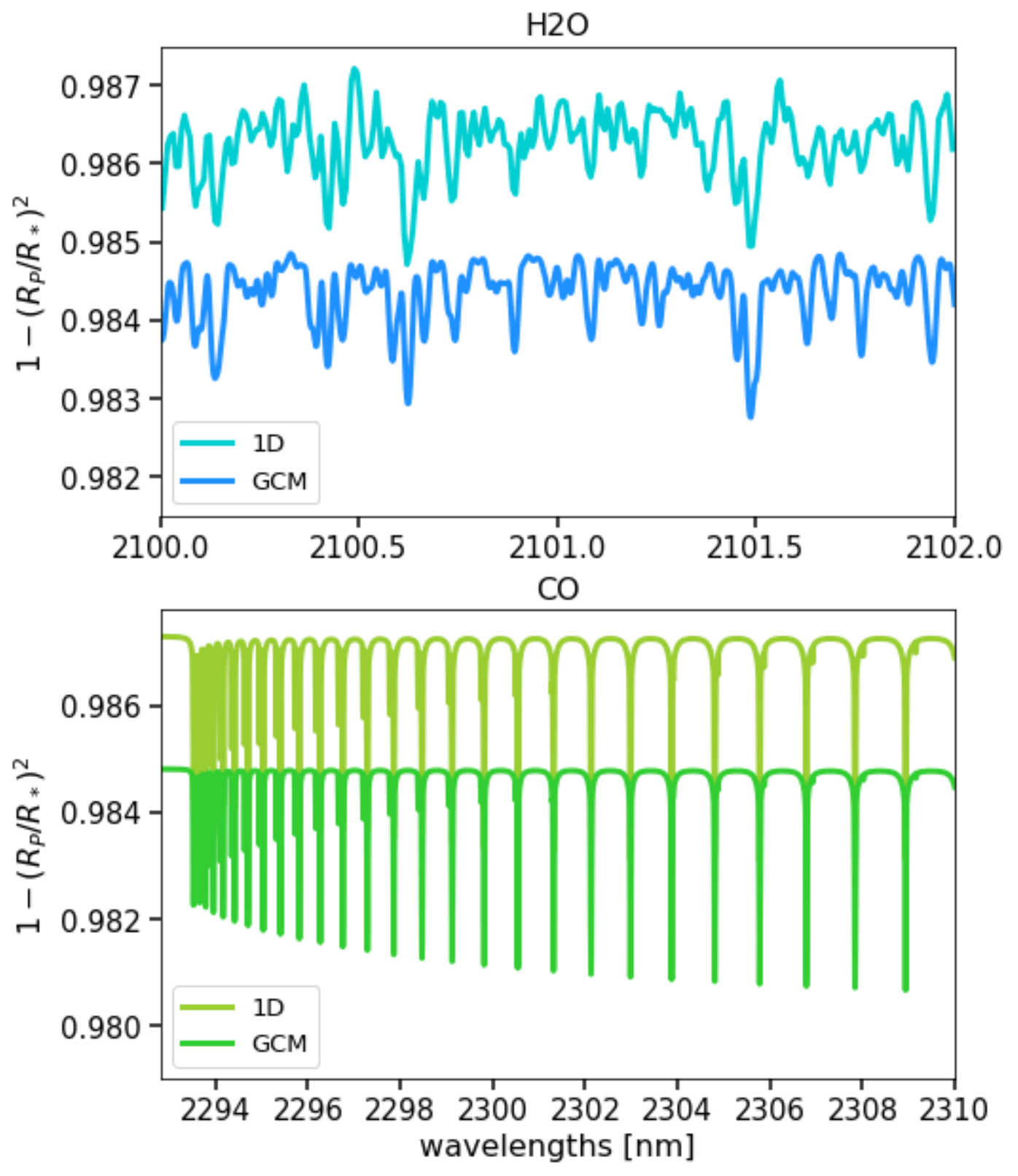}
      \end{tabular}
      \caption{\textit{Left:} Total of models used in this work (see Sec.~\ref{Sec:Methodology} for details), represented as the flux received from the star during transit, relative to the out-of-transit stellar flux, as function of  wavelengths. From left to right there are 1D models [panels \textcolor{gray}{(a)} and \textcolor{gray}{(b)}] and
      GCMs [panels \textcolor{gray}{(c)} and \textcolor{gray}{(d)}].
      In each plot, the gray line is the model with both  molecular species studied in this work, i.e., water and carbon monoxide. Only-H$_2$O [panels \textcolor{gray}{(a)},\textcolor{gray}{(c)}] and only-CO [panels \textcolor{gray}{(b)},\textcolor{gray}{(d)}] models are overplotted  with blue and green colors, respectively in the top and bottom panels. 
      Overall, GCMs (darker colors) have a deeper absorption respect with the 1D models (lighter colors) due to a different starting pressure at the bottom of the atmosphere. \textit{Right:} Zoom over a smaller wavelength range of only-H$_2$O (top panel) and only-CO (bottom panel) 1D and 3D models.}
        \label{Modelfullrange}
\end{figure*}

\begin{figure*}[!h]
   \centering
    \begin{tabular}{c}
    \includegraphics[width=0.5\hsize]{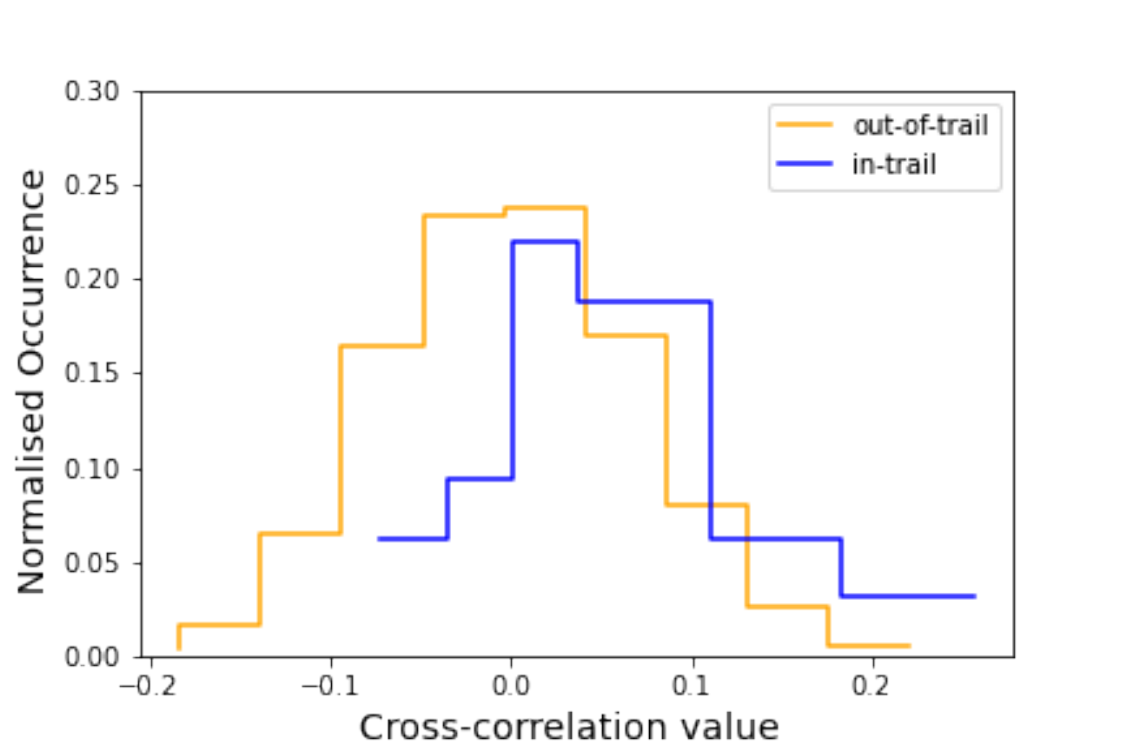}
    \includegraphics[width=0.5\hsize]{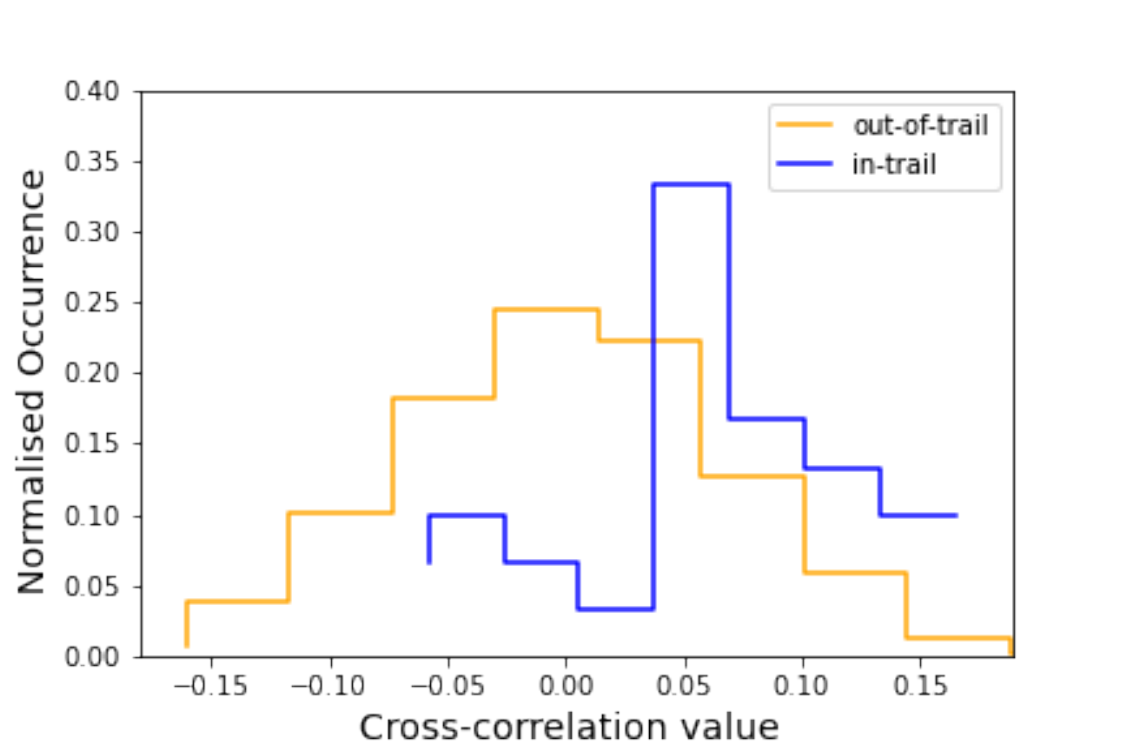}
      \end{tabular}
      \caption{Comparison between the distribution of cross-correlation values  outside (yellow solid line) and inside (blue solid line) the  radial velocity trail of WASP-20b obtained with the only-H$_2$O 1D model (left panel) and 3D model (right panel) used in this work. It is clear that the mean of the two distribution is not the same in both cases, being centered at 0 for the out-of-trail distribution and shifted towards positive values for the in-trail distribution.  A Welch t-test on the data rejects the hypothesis that the two distributions are drawn from the same parent distribution at the 3.1$\sigma$ of confidence level for 1D models and 4.1$\sigma$ for 3D models. }
        \label{in-out_trail}
\end{figure*}

	     




\end{document}